# A 28nm 1.80Mb/mm$^2$ Digital/Analog Hybrid SRAM-CIM Macro using 2D-Weighted Capacitor Array for Complex Number MAC Operations

Shota Konno[1,2], Che-Kai Liu[1], Sigang Ryu[1,3], Samuel Spetalnick[1], Arijit Raychowdhury[1]

[1]Georgia Institute of Technology, Atlanta, USA, [2]Asahi Kasei Microdevices Corporation, Yokohama, Japan, [3]Korea Aerospace University, Goyang-Si, Gyeonggi-do, South Korea

**Introduction**: Computing-in-memory (CIM) has attracted significant attention in recent years because of its energy-efficiency in edge inference applications. CIM has traditionally been developed for neural-network based image processing and has been demonstrated with real-numbers. However, many other sensor modalities such as RF, radar, biomedical, control data-processing etc. can leverage energy-efficient in-memory multiply–accumulate (MAC) [1-2] that can be enabled by CIM on complex numbers. In a conventional CIM on complex fields, the real and imaginary parts must be computed separately (including the cross-product terms), requiring separate control logic (with added area and power) for data orchestration. This can be achieved by: (1) duplicating the complex weights at 1.5x area to enable parallel computation of the partial products [3], or (2) sequentially computing at 2.2x latency. This paper presents a complex-CIM (C-CIM) macro prototype in 28nm CMOS featuring (1) a novel complex bit-cell that co-locates the real and imaginary parts thereby improving latency, area and computational density and reducing data-movement, (2) analog-CIM (ACIM) on the bitline using M7-M7 fringe at 48aF for LSBs at high area- and energy-efficiency, and (3) digital-CIM (DCIM) with counting logic for the MSB group that reduces area at high accuracy.

**Proposed Macro Architecture:** A complex CIM macro based on the 6T cell, is proposed to compute the real (Re) and imaginary (Im) parts in parallel, sharing the 6T CIM-SRAM for weights to improve latency and computational density (Fig.1). A 2D binary-weighted capacitor array (2D-Array) based on the product of input (operand 1) and weight (operand 2) implements a hybrid complex-CIM while eliminating the need for input DACs that are needed for conventional binary-weighted capacitor-based CIM [4-5]. This further improves accuracy at lower area overhead (Fig. 1) and eliminates the uncorrelated variations between RDACs and MOM CAPs. The prototype CIM macro demonstrates eight complex CIM-SRAM units and a total of 64kb (Fig. 1). Figure 2 shows the techniques applied to optimize the area of the proposed 2D-Array. In a naïve implementation, the CAP size will be significant, and especially in a 2D-Array, the effect would be squared (Fig. 2). In the signed magnitude format (SMF) [6], the MSB of the input and weight, I[7] and w[7], represent the sign bits, thus reducing the row and column of MSB (Fig. 2). LSB truncation has been reported to have a quantifiable but limited effect on ML workloads, and lightweight correction schemes have been further proposed for high-precision applications [7-10]. Leveraging these results, we calculate the contribution of each weight on the output accuracy and note that the top three MAC results account for half of the total contribution (Fig. 2). It's challenging to ensure matching accuracy for DACs with larger weights, as they become the dominant factor for the final output accuracy. This typically necessitates designing unit DACs with larger areas to reduce mismatch. By processing this group with DCIM, improved accuracy with minimal increase in digital circuit area can be achieved. Thus, in the proposed hybrid CIM architecture, DCIM is employed for only the top three upper-bit group, while ACIM is employed for the remaining bits. Therefore, the required ADC is 7b. Finally, the use of split-DAC effectively reduces the number of CAPs. An optimized 2D-Array is shown in Fig. 2. Figure 2 further shows the block diagram of the proposed hybrid CIM. In the DCIM, the + and - magnitude values are computed by the counting logic and adder tree in a time-multiplexed manner, and then subtracted to obtain a DCIM result in the range of +64 to -64. In the ACIM, the output of the multi-bit multiplier using pass transistor logic (PTL) is fed to the 2D-Array, the sum of the 16 units is calculated in the charge domain, and the resulting analog voltage is converted by the 7-bit SAR-ADC to produce the ACIM result. Post-digital adder combines the DCIM and ACIM results to obtain the final 8-bit CIM result. The design can be easily extended to higher bit-width if the application demands. Figure 3 shows the detailed implementation of 2D-Array ACIM. VREFCLK toggles direction according to SIGNCLK generated by Sign CKGEN to change the polarity of the analog voltage converted by the ADC. VREFSR uses 350mV, which allows the AND logic using pass transistors to be configured using only NMOS, contributing to area reduction. Since the CDAC of the ADC samples a fixed value of 0x40 when sampling, VREFAD uses 700mV, twice the VREFSR, to balance the charge range on the 2D-Array side. Figure 3 further shows the timing chart. The input and WLR vectors are updated at the start of the sampling phase ($\Phi_{SMP}$) and remain the same throughout the conversion. At the start of the next $\Phi_{SMP}$, the CIM results are updated. The Sign CKGEN generates the SGNCLK by inverting the polarity of the CNVCLK according to the sign bit. Figure 4 shows the floorplan of the complex CIM unit and die photo in 28-nm CMOS. The active area of the CIM macro is 0.0365 mm$^2$ and contains eight complex CIM units. Both 2D-Array and ADC CDAC are placed on top of the custom CIM-SRAM array for higher area-efficiency. This floorplan also improves capacitance matching between the 2D-Array and the ADC CDAC as they are adjacent to each other. The unit capacitor (UC) is 48aF, utilizing the parasitic capacitance of the M7-M7 fringe (Fig. 4). Since ACIM is used to process lower-order bits, we can use such a low UC, which contributes to overall area-savings. Compared to the minimum 2fF MOM provided by the foundry, the UC is one fortieth the size, with a very small unit area of 0.29um x 0.35um, thus enabling not only the 2D-Array design but also the SAR-ADC's CDAC to be placed above the SRAM array and DCIM, as shown in the cross-sectional view (Fig. 4). For the designed 48aF UC, a mismatch of 2.96% rms can be calculated based on foundry-provided minimum MOM CAP. With the proposed hybrid D/A architecture that relaxes the required ADC precision, the 7-bit binary CDAC, where the LSB is composed of 16C, results in a DNL of 0.33 LSB rms, meeting the linearity requirements. M5 is used for bottom node wiring and M6 for shielding. The first stage counting logic of the DCIM uses custom logic to reduce power and area (Fig. 9). The 6T-cell plus the counting logic in the DCIM results in better area efficiency compared to a conventional 6T plus local bit-wise MAC unit (e.g. with NOR gates [11]). Figure 4 shows a 64-word 6T-SRAM array for storing CIM weights. Double word-line (DWL) is used to avoid read-disturb. A low overhead write circuit with INV (with HVT-PMOS) and butterfly switch (SW) are used.

**Measurements and Results:** Figure 5 shows measurement results. The transfer function and INL (max INL at zero crossing) are measured by sweeping the input from negative full scale (FS) to positive FS while weights are fixed at negative FS, i.e. -127. Good linearity was obtained without calibration, with almost no gain error. The measured RMS error of the complex MAC (C-MAC) operation under uniform input conditions without considering sparsity is 0.435% rms, the lowest level compared to previous CIM prototypes [3-6, 12-14] (Fig. 6). The ACIM power dominates because of the low DCIM computation enabled by the topology. An energy efficiency of 35.0 TOPS/W is measured. Finally, the end-to-end function of the C-MAC operation is shown, with weights fixed at -127 and Re/Im values of input vectors swept across the full range. The memory density of 1.80 Mb/mm$^2$ is achieved, which is 2x compared to previous 6T-SRAM prototypes [12-13], and comparable to eNVM-based CIM [14]. Figure 7 shows the die microphoto and key operating parameters.

**Acknowledgments:** This work was supported in part by CoCoSys, one of seven centers in JUMP 2.0, a Semiconductor Research Corporation (SRC) program sponsored by DARPA. Correspondence and comments to Che-Kai Liu <che-kai@gatech.edu>.

## Prior Art

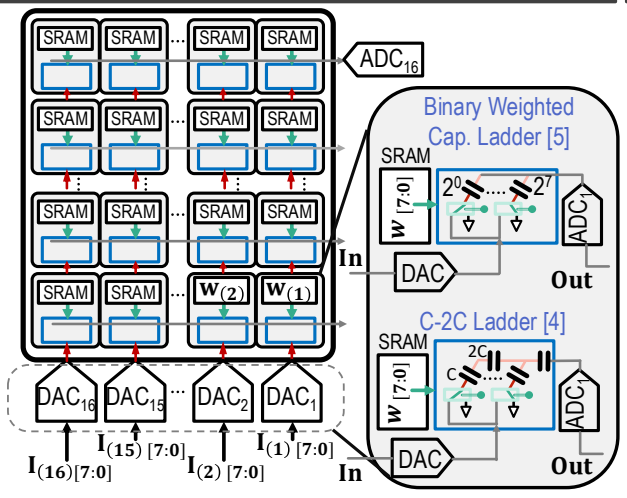

## Challenges and Contributions

| Challenge | Importance | Prior Art | Contribution |
|---|---|---|---|
| Multi-bit CIM Support | System Scalablity | C-2C Ladder [4] | 2D-Weighted Capacitor Array |
| CIM Accuracy | Data-Dependent MAC Errors | Digital / Analog | Hybrid A/D Architecture |
| Macro Density | CIM & Data Movement Tradoff | Low Density | 48aF-unit C Arr. / DAC-less Input |
| Robust Complex MAC Operation | DSP & ML Application Support | NR / Analog [3] | Measured Robust 0.435% rms C-MAC Operation |

## Proposed Capacitor-based CIM Macro & System Overview

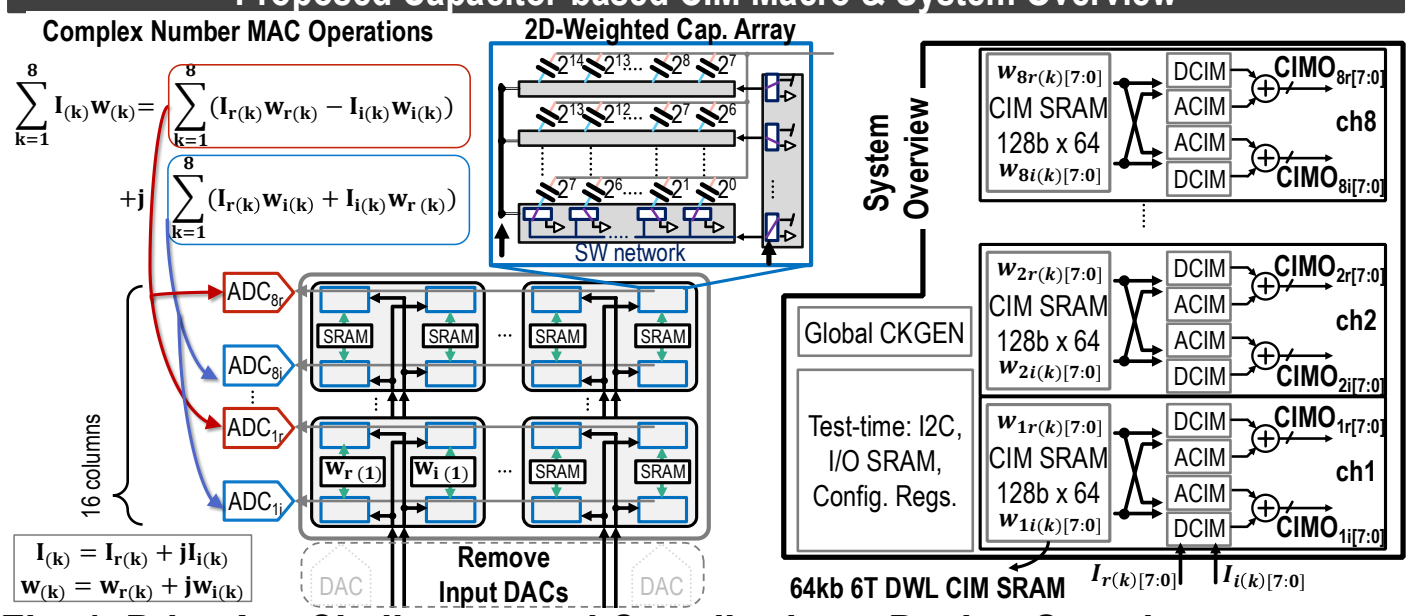

Fig. 1. Prior Art. Challenges and Contributions. Design Overview.

## 2D-Weighted Capacitor Array – Data Format & Benchmarks

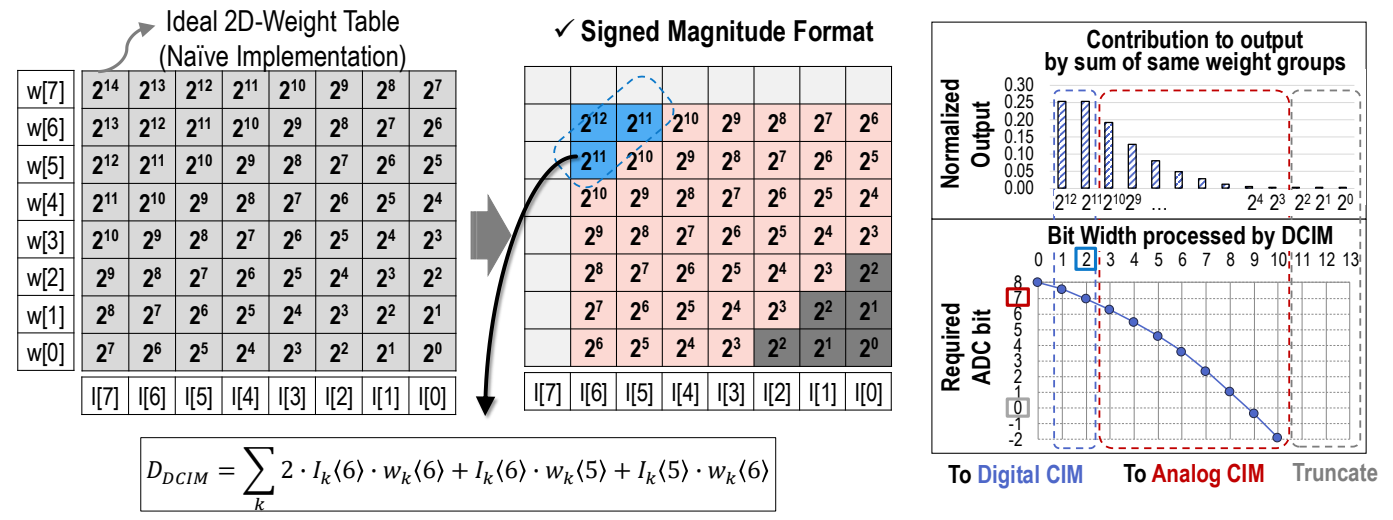

$$D_{DCIM} = \sum_{k} 2 \cdot I_k\langle 6\rangle \cdot w_k\langle 6\rangle + I_k\langle 6\rangle \cdot w_k\langle 5\rangle + I_k\langle 5\rangle \cdot w_k\langle 6\rangle$$

## Digital /Analog Hybrid CIM Architecture

- ✓ Digital/Analog Hybrid CIM Arch.
- ✓ Ignore LSBs
- ✓ Splitted C-DAC

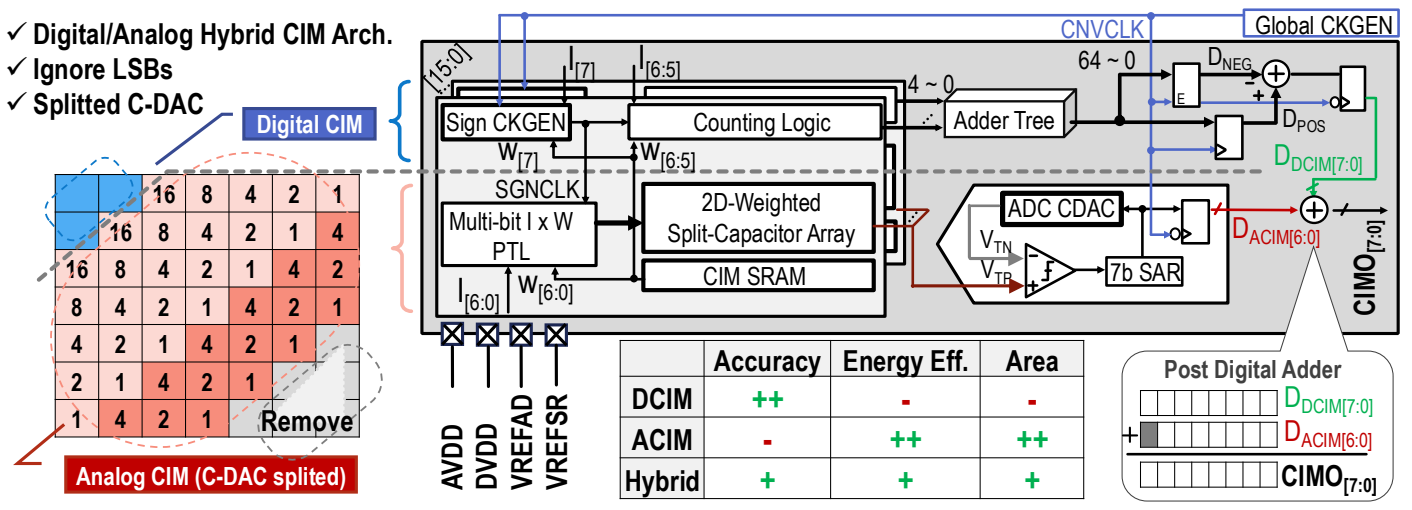

| | Accuracy | Energy Eff. | Area |
|---|---|---|---|
| DCIM | ++ | - | - |
| ACIM | - | ++ | ++ |
| Hybrid | + | + | + |

Fig. 2. SMF and the Proposed Hybrid D/A CIM Architecture.

## Analog CIM Implementation

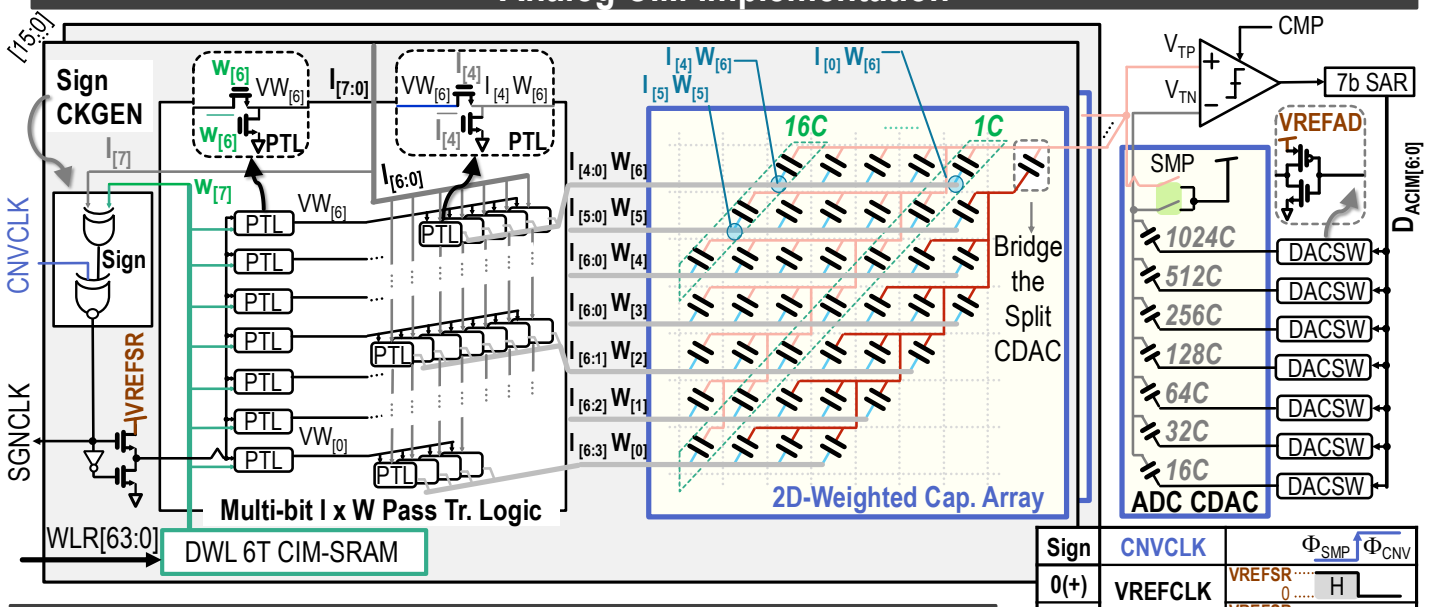

## Counting Logic on 1st Stage Digital CIM

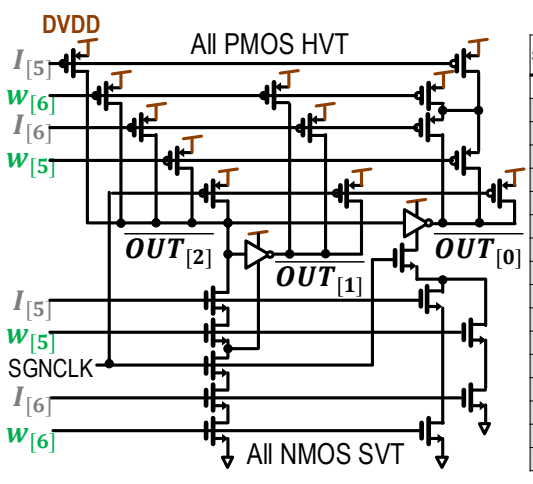

| SGNCLK | I[6] | I[5] | w[6] | w[5] | OUT [2] | OUT [1] | OUT [0] | Decimal |
|---|---|---|---|---|---|---|---|---|
| L | * | * | * | * | L | L | L | 0 |
| H | L | L | L | L | L | L | L | 0 |
| H | L | L | H | L | L | L | L | 0 |
| H | L | H | L | L | L | L | L | 0 |
| H | L | H | H | L | L | L | H | 1 |
| H | L | L | L | H | L | L | L | 0 |
| H | L | L | H | H | L | L | L | 0 |
| H | L | H | L | H | L | L | L | 0 |
| H | L | H | H | H | L | L | H | 1 |
| H | H | L | L | L | L | L | L | 0 |
| H | H | L | H | L | L | H | L | 2 |
| H | H | H | L | L | L | L | L | 0 |
| H | H | H | H | L | L | H | H | 3 |
| H | H | L | L | H | L | L | H | 1 |
| H | H | L | H | H | L | H | H | 3 |
| H | H | H | L | H | L | L | H | 1 |
| H | H | H | H | H | H | L | L | 4 |

## Timing Chart

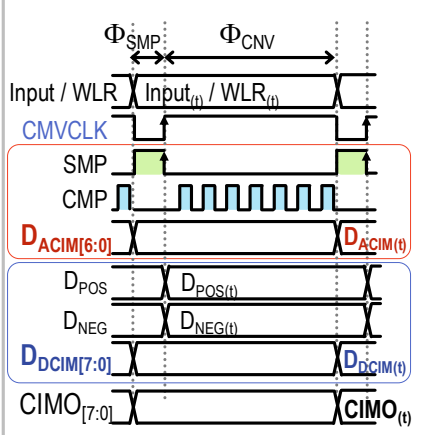

Fig. 3. 2D-Array ACIM. Digital Counting Logic. Timing Chart.

## Complex (Real and Imaginary) CIM Unit and Capacitor Array Floorplan

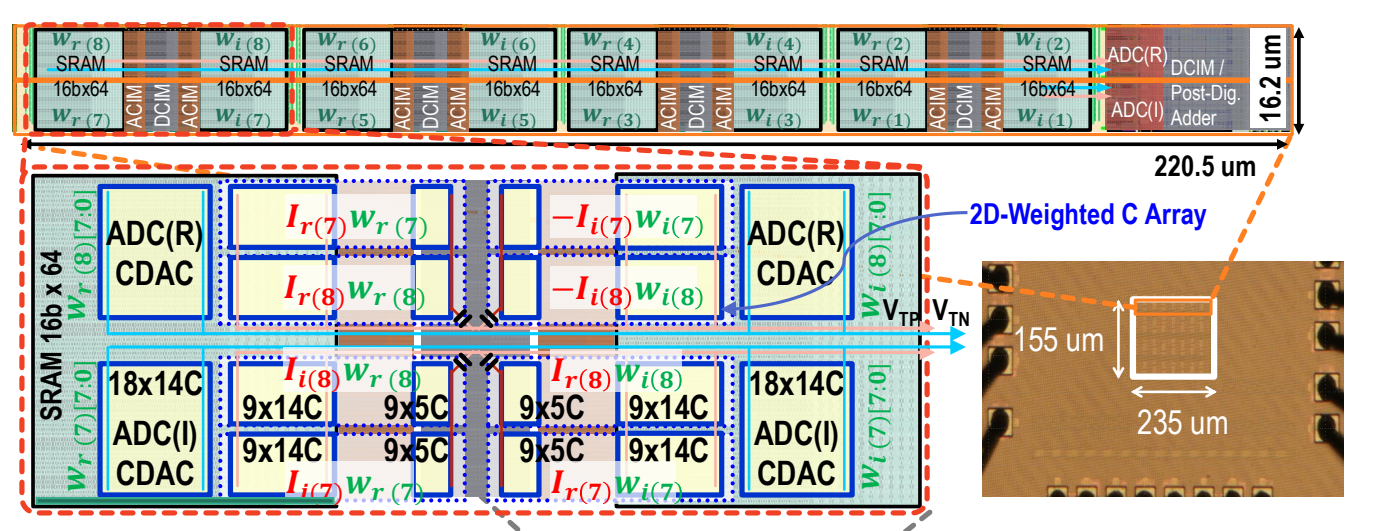

## Double WL 6T-SRAM Array

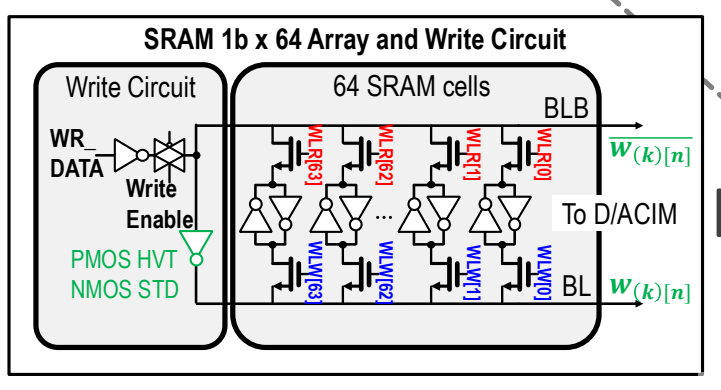

| Mode | Write Enable | WLR [k] | WLW [k] |
|---|---|---|---|
| Read (CIM operation) | L | [k] on | All off |
| Write | H | [k] on | [k] on |

## Cross-sectional View

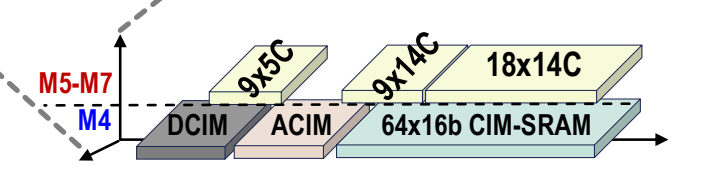

## Unit Capacitor Layout

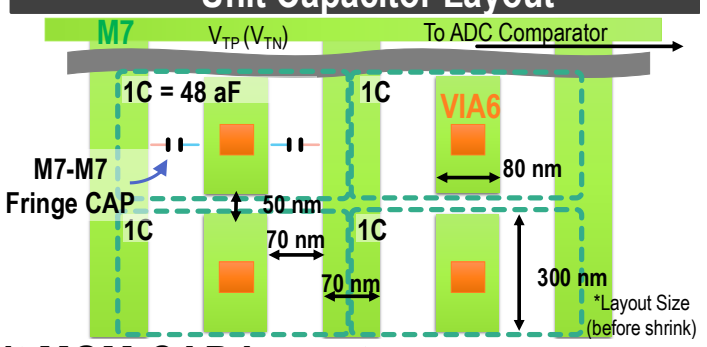

Fig. 4. Macro Floorplan. CIM-SRAM. Unit MOM-CAP Layout.

## Integral Non-linearity (INL)

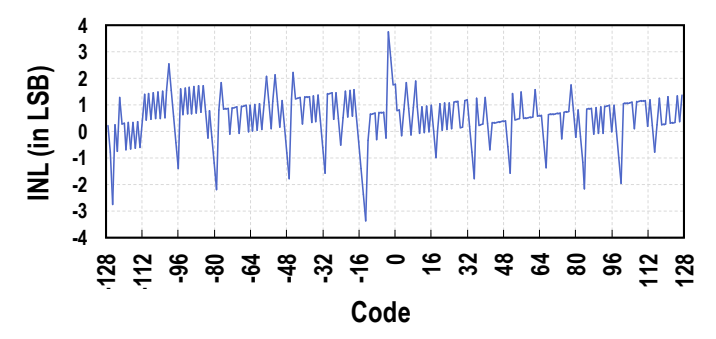

## MAC RMS error

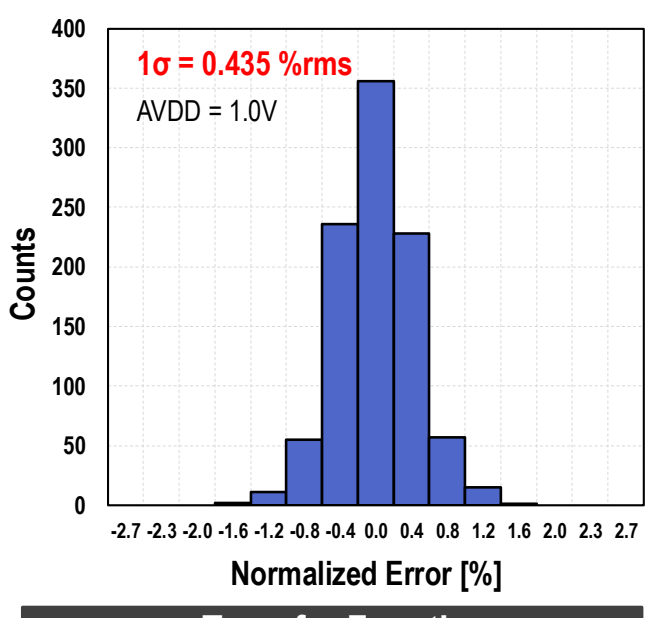

## Power Breakdown

| Name | Supply for | Voltage | Power |
|---|---|---|---|
| AVDD | 6T CIM-SRAM & ACIM | 900 mV | 680 uW |
| DVDD | DCIM | 750 mV | 132 uW |
| VREFAD | ADC C-DAC | 600 mV | 173 uW |
| VREFSR | 2D-Weighted Cap. Array | 300 mV | 39 uW |

## Transfer Function

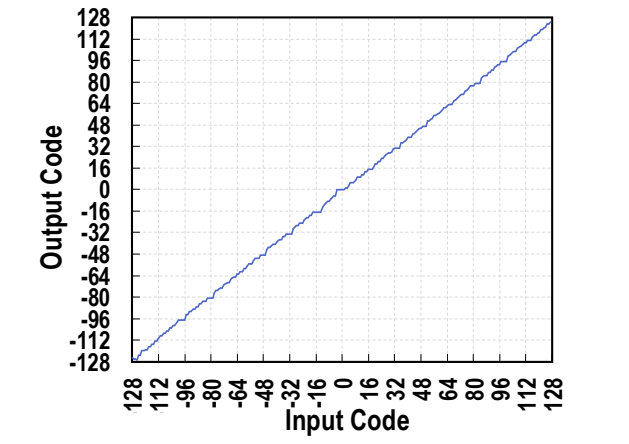

## Complex Number End-to-End Function

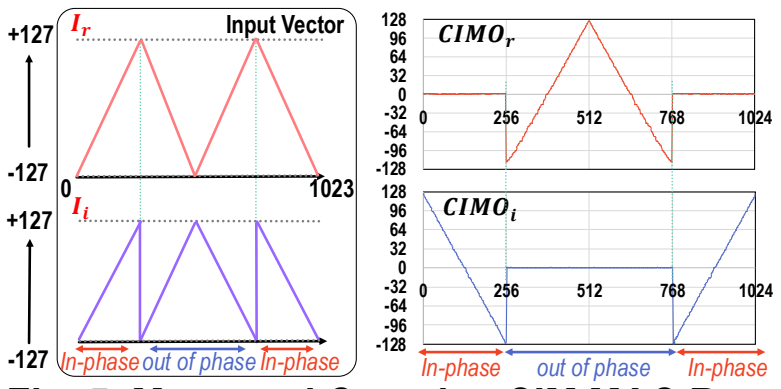

Fig. 5. Measured Complex-CIM MAC Results.

| | This work | | ASSCC'23 [3] Ma *et al.* | JSSC'23 [4] Wang *et al.* | JSSC'24 [6] Kim *et al.* | ISSCC'22 [12] Yan *et al.* | JSSC'23 [13] Wu *et al.* | ISSCC'23 [5] Hsieh *et al.* | VLSI'23 [14] Spetalnick *et al.* |
|---|---|---|---|---|---|---|---|---|---|
| Tech. [nm] | 28 | | 40 | 22 | 28 | 28 | 28 | 12 | 40 |
| CIM Architecture | Hybrid | | Analog | Analog | Analog | Digital | Time-Domain | Analog | Analog |
| ACIM MAC Architecture | 2D-W CDAC w. 48aF UC | | Current | C-2C CDAC | MOM Capacitor | - | - | Binary CDAC | Current |
| Cell Type | 6T SRAM | | 8T SRAM | 9T SRAM | 3T-2C eDRAM | 6T SRAM | 6T SRAM | 9T SRAM | 1T-1R RRAM |
| Area [mm²] | 0.0365 | | 17.94 | 0.25 | 0.103 | 0.030 | 1.22 (4) | N/A | 0.0263 |
| CIM memory bit | 64kb | | 512kb | 128kb | 100kb | 32kb | 1Mb | 128kb | 64kb |
| Memory Density [Mb/mm²] | 1.80 | | 0.029 | 0.512 | 0.95 | 1.067 | 0.85 (4) | N/A | 2.38 |
| Precision (Input : Weight) | 8b:8b | | BF16-FP32: BF16-FP32 | 8b:8b | 4b:5b | 1b-8b:1b/4b/8b | 8b:8b | 8b:8b | 1b:1b |
| RMS error [%rms] | 0.435 | | N/A | 0.89 / 0.46 (pre-distorted) | 0.80 (3) | 0 (Digital) | N/A | N/A | 2.8 (5) |
| Throughput [GOPS] | 35.8 (1) | 46.6 (2) | 5.08 GFLOPS | 600 – 1000 | 19.5 (4b-5b) | N/A | 1050 - 1241.2 | 102.4 | 33 - 266 (1b-1b) |
| Energy Efficiency [TOPS/W] | 35.0 (1) | 25.8 (2) | 4.74 TFLOPS/W | 15.5 - 32.2 | 115.8 (4b:5b) | 27.38 | 28.66 - 39.31 | 70.85 - 86.27 | 9.81 - 75.17 (1b-1b) |
| Compute Density [TOPS/mm²] | 0.98 (1) | 1.28 (2) | N/A | 2.4 - 4.0 | 2.03 (4b:5b) | N/A | 0.86 - 1.02 | N/A | 0.87 - 7.0 (1b-1b) |
| C-MAC Support | ✓ | | ✓ | ✗ | ✗ | ✗ | ✗ | ✗ | ✗ |

1. AVDD = 0.9V 2. AVDD = 1.0V 3. Estimated from RMS err. (Outlier) 4. Estimated from Throughput and Area Efficiency 5. Estimated from MAC result RMS error (CIM Mode = 16)

Fig. 6. Comparison Table.

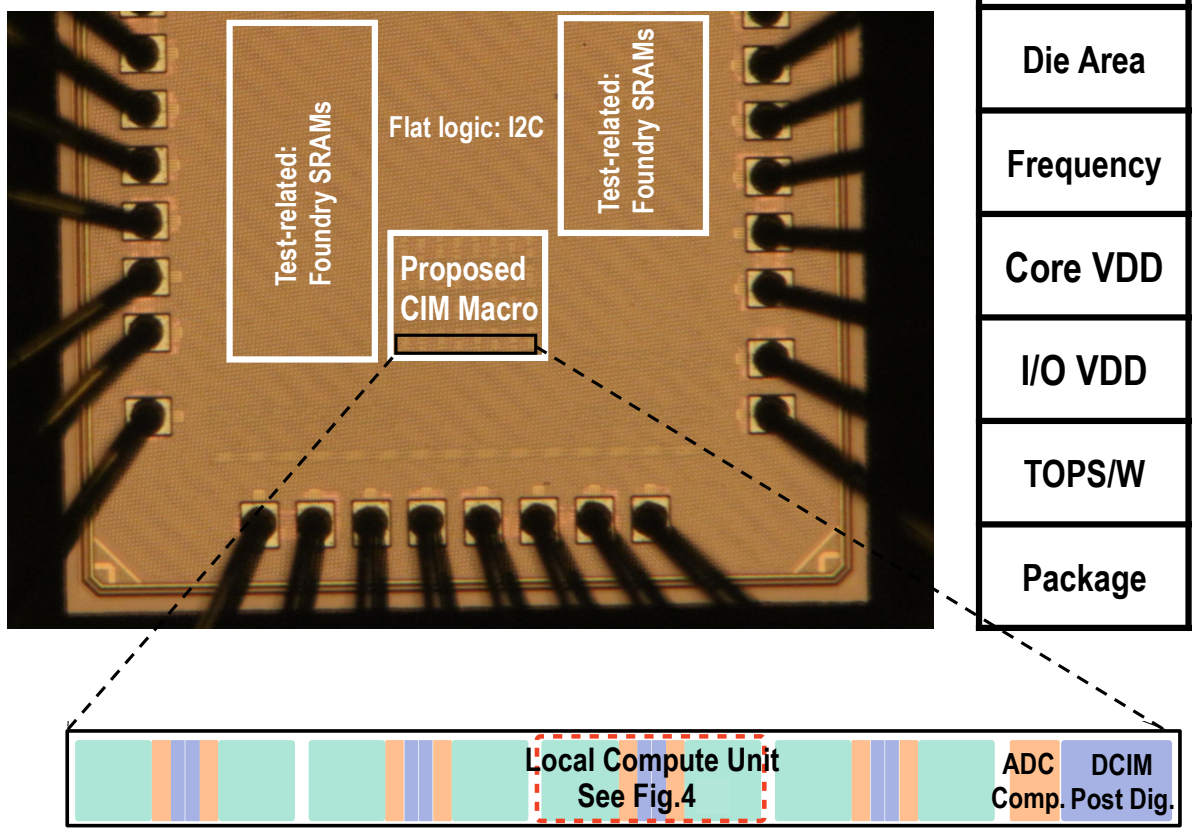

| Technology | N28 HPC+RF |
|---|---|
| Die Area | $1mm^2$ |
| Frequency | 91 MHz |
| Core VDD | 0.9 / 0.75 / 0.6 / 0.3 V |
| I/O VDD | 1.8 V |
| TOPS/W | 35.0 |
| Package | QFN 32 |

Fig. 7. Die Microphoto, design parameters and key operating parameters

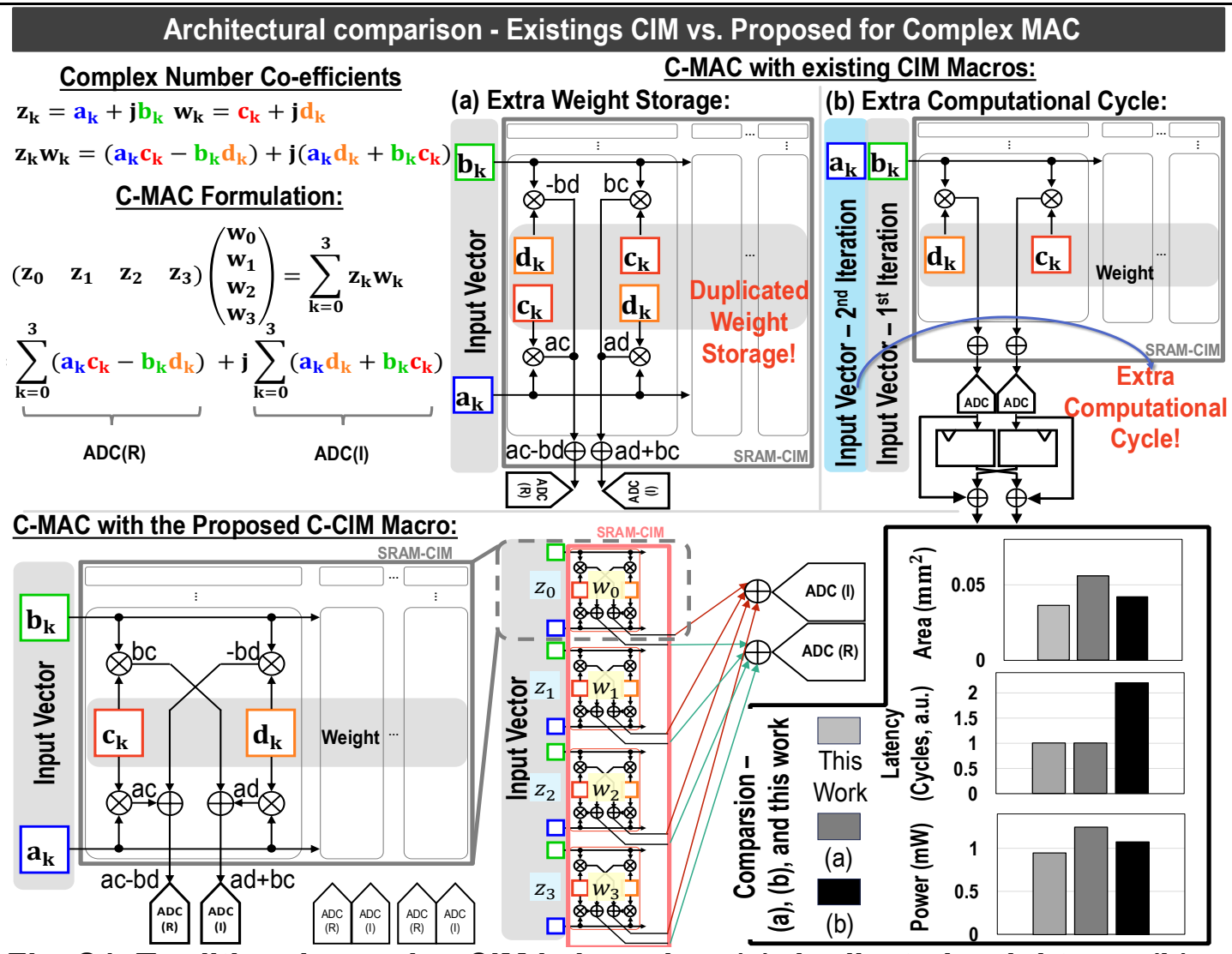

Fig. S1. Traditional complex-CIM is based on (a) duplicated weights, or (b) extra computational cycles and extra control logic. This work shows lower area (35%), latency (54%) and power (24%) vs. the best of (a) or (b).

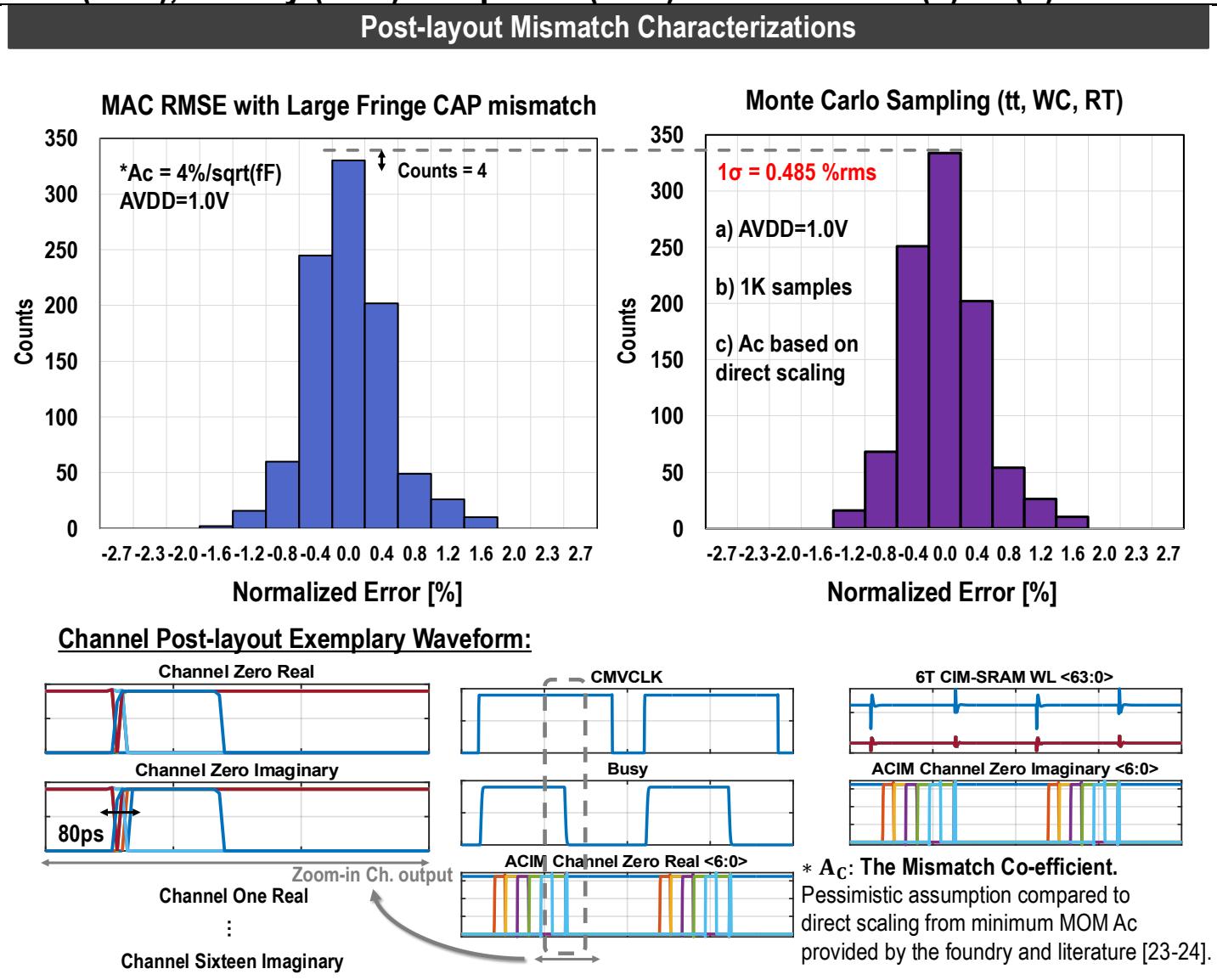

Fig. S2. Simulated Monte Carlo analysis of RMS error at target CAP mismatch illustrating the design viability.

## Additional References:

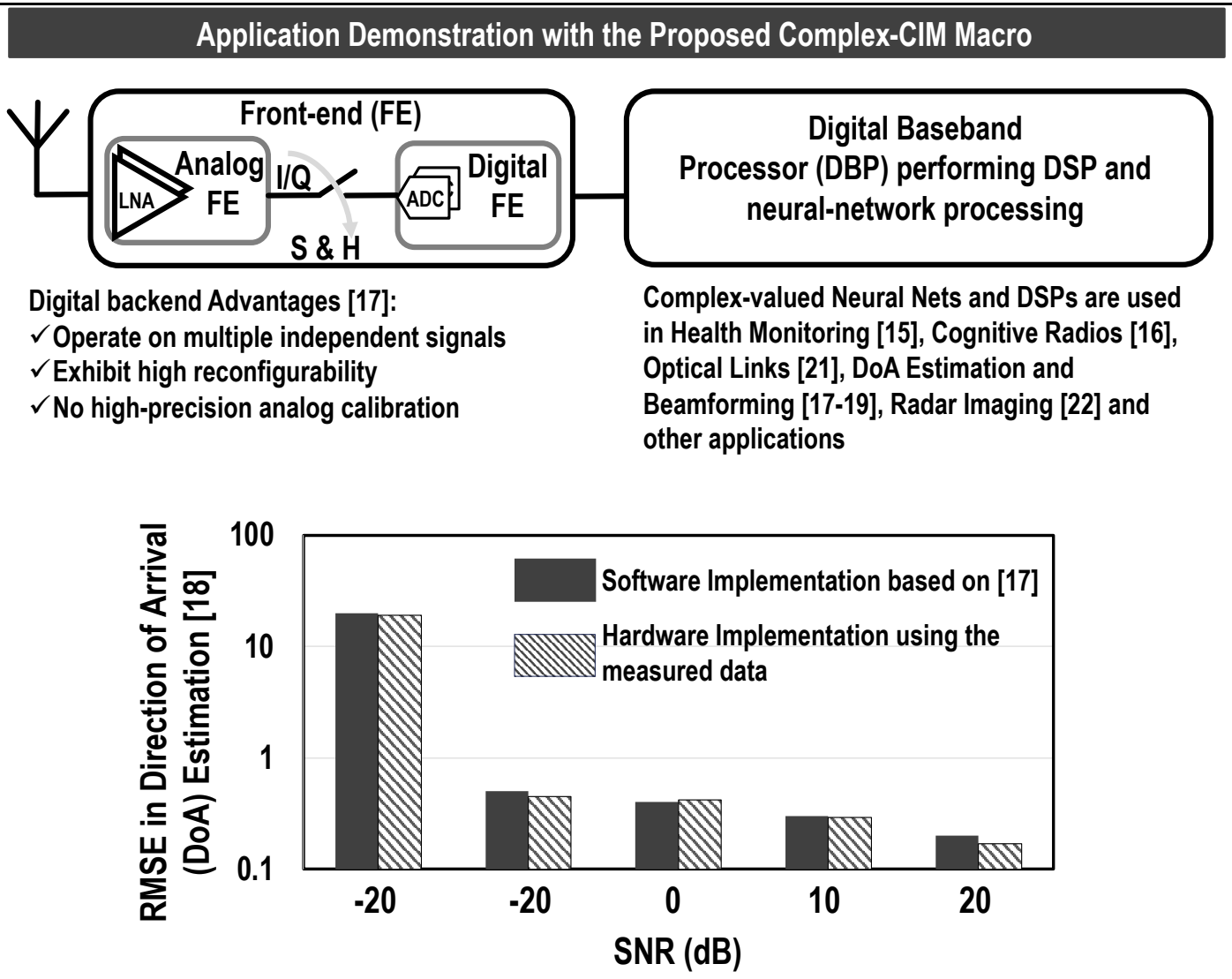

Fig. S3. Complex-CIM finds many applications in signal-/neural-processing. Typical application in DOA Estimation [18] shows less than 4% RMSE error with the proposed design compared to software implementation.